\documentclass{iaus}
\usepackage{graphicx}

\title[Apsidal Precession in Double White Dwarf Systems ] %% give here short title %%
{Constraining White Dwarf Masses Via  Apsidal Precession in Eccentric Double White Dwarf Binaries. }

\author[Valsecchi et al.]   %% give here short author list %%
{Francesca Valsecchi$^1$, Will M. Farr$^1$, Bart Willems$^1$, Christopher J. Deloye$^1$ \& Vicky Kalogera$^1$}

\affiliation{$^1$Center for Interdisciplinary Exploration and Research in Astrophysics (CIERA), and Northwestern University, Department of Physics and Astronomy, 2145 Sheridan Road, Evanston, IL 60208, USA.}

\pubyear{2011}
\volume{281}  %% insert here IAU Symposium No.
\pagerange{119--126}
\jname{Binary Paths to Type Ia Supernovae Explosions}
\editors{A.C. Editor, B.D. Editor \& C.E. Editor, eds.}
\begin{document}

\maketitle

\begin{abstract}

Galactic short period double white dwarfs (DWD) are guaranteed gravitational wave (GW) sources for the next generation of space-based interferometers sensitive to low-frequency GWs ($10^{-4}$- 1$\,$Hz). Here we investigate the possibility of constraining the white dwarf (WD) properties through measurements of apsidal precession in eccentric binaries. We analyze the general relativistic (GR), tidal, and rotational contributions to apsidal precession by using detailed He WD models. We find that apsidal precession can lead to a detectable shift in the emitted GW signal, the effect being stronger (weaker) for binaries hosting hot (cool) WDs. We find that in hot (cool) DWDs tides dominate the precession at orbital frequencies above $\sim$0.01 mHz ($\sim$1 mHz). Analyzing the apsidal precession of these sources only accounting for GR would potentially lead to an extreme overestimate of the component masses. Finally, we derive a relation that ties the radius and apsidal precession constant of cool WD components to their masses, therefore allowing tides to be used as an additional mass measurement tool.

%% add here a maximum of 10 keywords, to be taken form the file <Keywords.txt>
\end{abstract}

\firstsection % if your document starts with a section,
              % remove some space above using this command.
\section{Introduction}
The Laser Interferometer Space Antenna (LISA) is expected to provide the largest observational sample of short period DWDs (\cite{NelemansEtAl2001a}, \cite[2001b]{NelemansEtAl2001b}, \cite[2004]{NelemansEtAl2004}, \cite{LiuEtAl2010,RuiterEtAl2010}). 
Recent theoretical calculations by  \cite[Willems et al. (2007)]{Willems2007} and \cite[Thompson (2010)]{Thompson2010} predicted a population of eccentric DWDs. \cite[Willems et al. (2007)]{Willems2007} studied eccentric binaries formed through dynamical interactions in globular clusters, while \cite[Thompson (2010)]{Thompson2010} studied eccentric DWDs as products of hierarchical triple systems, where the tertiary induces Kozai oscillations in the inner binary (see also \cite{Gould2011} for a discussion of the implications of such a population).
Our focus here lies on these eccentric binaries, where measurements of apsidal precession could be used to constrain some of the WD properties. For this purpose, we use detailed He WD models with masses between 0.1-0.3$\, M_\odot$ (\cite{Deloye2007}) to investigate apsidal precession in eccentric binaries.
In what follows, we consider a binary hosting two stars with masses $M_{1,2}$, and radii $R_{1,2}$, uniformly rotating with angular velocities $\Omega_{1,2}$. We take the axes of rotation to be orthogonal to the orbital plane. Let $\gamma$ be the argument of the periastron,  $P$ and $a$ the orbital period and semi-major axis, respectively, and $e$ the orbital eccentricity. We consider orbital and rotational periods long compared to the free harmonic periods of the component stars (\cite{Cowling1938, Sterne1939, SmeyersWillems2001}), so that we can neglect resonances between the tidal forcing angular frequencies and the eigenfrequencies of WD free oscillation modes, that is, we work in the static-tide limit.

\section{Apsidal Precession and Driving Mechanisms}
Apsidal precession occurs when the binary's gravitational potential is perturbed from a pure Keplerian potential. This perturbation can be due, for example, to tides, rotation, and GR.  
The contribution to the apsidal precession due to the quadrupole tides raised in star $i$ by the companion is (\cite{Sterne1939}):
\begin{equation}
\dot\gamma_{Tid, i} = \frac{30\pi}{P}\left(\frac{R_i}{a}\right)^5\frac{M_{3-i}}{M_i}\frac{1+\frac{3}{2}e^2+\frac{1}{8}e^4}{(1-e^2)^5}k_i,
\label{eq:gammaDotTid}
\end{equation}
where $k_i$ is the so-called apsidal precession constant for star $i$, and it measures the star's central concentration. 
We refer to \cite[Valsecchi et al. (2011)]{Valsecchietal2011} for details on how to calculate  $k_i$ from detailed WD models.

The apsidal precession due to the rotational distortion of the star caused by centrifugal forces takes the form (e.g. \cite{Sterne1939}):
\begin{equation}
\dot\gamma_{Rot, i} = \frac{2\pi}{P}\left(\frac{R_i}{a}\right)^5\frac{M_1+M_2}{M_i}\frac{(\Omega_i/\Omega)^2}{(1-e^2)^2}k_i,
\label{eq:gammaDotRot}
\end{equation}
where $\Omega = 2\pi/P$ is the mean motion.
Finally, the GR contribution to apsidal precession at the leading quadrupole order is given by  (\cite{LeviCivita1937}): 
\begin{equation}
\dot\gamma_{GR} = \left(\frac{2\pi}{P}\right)^{5/3}\frac{3G^{2/3}}{c^2}\frac{(M_1+M_2)^{2/3}}{(1-e^2)},
\label{eq:gammaDotGR}
\end{equation}
where $G$ and $c$ are the gravitational constant and the speed of light, respectively. The total apsidal precession rate ($\dot{\gamma}$) is the sum of the three contributions above considering both components. We note that all contributions to $\dot{\gamma}$ depend on $P$, $e$ and $M_i$ (with $\dot{\gamma{_{GR}}}$ being dependent upon the total system mass), while only $\dot\gamma_{Rot, i}$ and $\dot\gamma_{Tid, i}$ depend on $R_i$ and $k_i$.

\section{Apsidal Precession in a Binary Hosting Two 0.3$M_{\odot}$ Components}\label{calculatedApsidalPrecession}
\begin{figure} [h]
   \centering
\includegraphics[width=7.8cm]{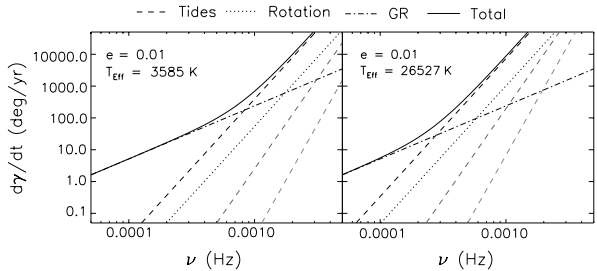}
   \caption{Apsidal precession rates for cool (\textit{left}) and hot (\textit{right}) DWDs with masses $M_1 = M_2 = 0.3\, M_\odot$, and orbital eccentricity $e = $0.01. The tidal and rotational rates include the contribution of both stars. The rotational contribution is calculated assuming synchronization at periastron. For cool WDs $\dot\gamma_{Tid, i}$ and $\dot\gamma_{Rot, i}$ decrease because of their dependence on $R_i^5$.}
   \label{fig:periPrec}
    \end{figure}

In Fig. \ref{fig:periPrec} we show an example of tidal, rotational, and GR apsidal precession rates calculated with our models, as a function of the orbital frequency $\nu = 1/P$. In general, GR dominates $\dot{\gamma}$ at small $\nu$, while tides and rotation become more significant as $\nu$ increases. The currently observed sample of rotation rates in DWDs suggests that apsidal precession is tidally-induced at frequencies where GR is no longer important (see \cite{Valsecchietal2011} for details).  Apsidal precession can be significant for both hot and cool DWDs. For cool components at $\nu \simeq 1\,$mHz, and $e = 0.01$, apsidal precession is expected to shift the GW signal by 3$\pi$-4$\pi$ over 1$\,$yr mission, which is detectable in proposed space based interferometers (see \cite{WillemsEtAl2008} for a detectability study). The shift increases by more than a factor of 10 for hot WDs, and at $\nu\simeq 1\,$mHz tides dominate. Note that, due to the dependency of $\dot\gamma_{Tid, i}$ on $R_i^5$, binaries hosting hot (and therefore bigger) WDs have $\dot{\gamma_{Tid}}$ systematically higher than cool (and contracted) WDs by a factor of $\sim 20$.

With lower component masses, $\dot\gamma_{GR, i}$ decreases, while $\dot\gamma_{Tid, i}$ and $\dot\gamma_{Rot, i}$ increase. The larger radii of the lower mass models act to increase the tidal and rotational contributions relative to GR, therefore making detection even more favorable. 
Note that theoretical calculations from \cite[Willems et al. (2007)]{Willems2007} predicted orbital eccentricity distributions (affected by post-formation GW emission) that typically peak at $e\simeq$0.1, with a non- negligible fraction of systems having $e$ up to $\simeq$0.7. Increasing $e$ would lead to much higher rates of apsidal precession (see \cite{Valsecchietal2011}). 
\section{Tidal Contributions to Apsidal Precession}\label{Sect:tidesPeriPrec}
Given the dependence of $\dot\gamma_{GR}$ only on $P$, $e$, and $M_i$, if apsidal precession is detected in binaries where GR is the leading mechanism, it can be used to place constraints on the total system mass. Here we show that $\dot\gamma_{Tid, i}$ can be used in a similar fashion considering the evolution of the term $k_i R_i^5$ entering Eq. (\ref{eq:gammaDotTid}) as the WD cools.
\begin{figure} [h]
   \centering
\includegraphics[width=5.8cm]{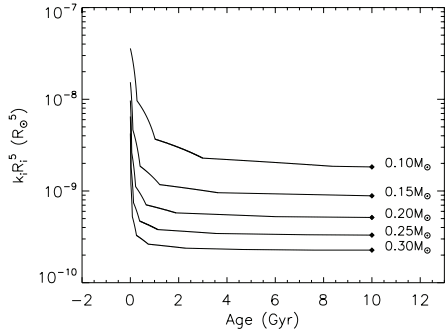}
   \caption{Evolution of $k_iR_i^5$ as a function of the WD's age. Filled squares denote plateau values. The maximum age considered for each model is 10$\,$Gyr.}
   \label{fig:k2R5_age}
    \end{figure}
As Figure \ref{fig:k2R5_age}  shows, during most of the WD's lifetime $k_i R_i^5$ assumes a plateau value which is unique for each mass. Consequently Eq. (\ref{eq:gammaDotTid}) effectively depends only on $P$, $e$, and $M_i$ for much of the WD lifetime. A fit of the plateau values of $k_i R_i^5$ as a function of the WD mass yields $k_iR_i^5 = -0.632+0.370\cdot M_i^{ -1.709}$, where $M_i$ is the mass in solar units, and $k_iR_i^5$ is in $10^{-10} R_\odot^5$ .
Therefore, for any DWD for which apsidal precession is detectable, $\dot\gamma$ yields constraints on a particular combination of the masses in the
system, which we call the "apsidal mass function". 
\section{Misclassification of the Sources}\label{Sect:SourcesMisclass}
\begin{figure} [h]
   \centering
\includegraphics[width=8.8cm]{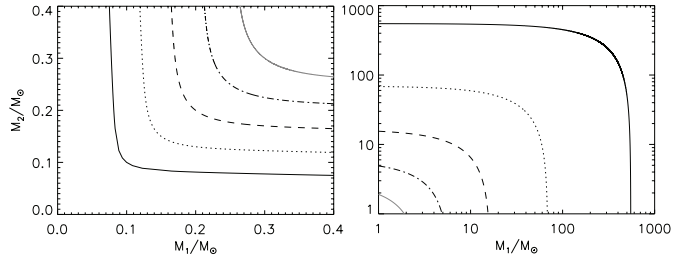}
   \caption{Lines of constant apsidal precession rate calculated for cool DWDs with $\nu=1\, mHz$, and $e = 0.01$. The left panel depicts lines of constant total $\dot{\gamma}$ calculated for different combinations of components' masses accounting for GR, rotation and tides. The right plot shows, for these same $\dot{\gamma}$, the masses that would be inferred solely accounting for GR. The various lines correspond to $\dot\gamma$ in deg/yr of $\simeq$ 22570 (black-solid), $\simeq$ 5700 (dotted), $\simeq$ 2200 (dashed), $\simeq$ 1100 (dot-dashed), $\simeq$ 680 (grey-solid).}
   \label{fig:DetectionError}
    \end{figure}

Analyzing an apsidal precession detection only accounting for the GR contribution, could potentially lead to a dramatic overestimate of the components masses. As shown in Figure \ref{fig:DetectionError}, an eccentric DWD could be misclassified as a binary black hole or neutron star if the total apsidal precession is mis-attributed to GR alone.

\section{Summary and Conclusions}\label{Sect:conclusions}
We use detailed He WD models to investigate apsidal precession due to tides, rotation and GR in eccentric DWDs.
We find that apsidal motion can lead to a significant shift in the emitted GW signal, the effect being stronger for binaries with hot WDs. In general, GR dominates the precession at small orbital frequencies $\nu$, while $\dot{\gamma}$ is driven by tides at larger frequencies, where GR is no longer important.
We investigate the astrophysical information that can be extracted from a measured $\dot{\gamma}$. We show that, as the WD ages, $k_iR_i^5$ reaches a plateau value that depends only on the WD mass. This behavior allows us to simplify the form of Eq. (\ref{eq:gammaDotTid}), and to use  $\dot\gamma_{Tid, i}$ to place constraints on the apsidal mass function also for short-period binaries, extending the possibility of mass measurement to the entire LISA frequency domain. We also find that the component masses that would be inferred from the GR contribution alone could be overestimated by orders of magnitude.
We conclude that the inclusion of tides is necessary for the proper analysis of detected apsidal motion rates of DWDs in the LISA band.

\begin{acknowledgements}
      This work was supported by NASA Award NNX09AJ56G.
\end{acknowledgements}

\end{document}